\documentclass[twoside,english]{iopart}
\usepackage[T1]{fontenc}
\usepackage[latin9]{inputenc}
\usepackage{geometry}
\usepackage{amstext}
\usepackage{graphicx}

\makeatletter

\providecommand{\tabularnewline}{\\}

\usepackage{iopams}
\usepackage{setstack}

\usepackage[numbers, sort&compress]{natbib}

\newcommand{\eqref}[1]{(\ref{#1})}

\makeatother

\usepackage{babel}
\begin{document}

\title{Cooperative effects of lattice and spin-orbit coupling on the electronic
structure of  orthorhombic SrIrO$_{3}$ }

\author{Vijeta Singh$^{1,2}$ and J J Pulikkotil$^{1,2,3}$ }

\address{$^{1}$Quantum Phenomena \& Applications Division, CSIR-National
Physical Laboratory, New Delhi 110012\\
$^{2}$Academy of Scientific \& Innovative Research (AcSIR), CSIR-National
Physical Laboratory, New Delhi 110012\\
$^{3}$Computation and Network Facility, CSIR-National Physical Laboratory,
New Delhi 110012 \\
}

\ead{vijetasingh@nplindia.org, jiji@nplindia.org}
\begin{abstract}
Orthorhombic SrIrO$_{3}$ subjected to strain show tunable transport
properties. With underlying symmetry remaining invariant, these properties
are associated with IrO$_{6}$ octahedral tilting. Adopting to first-principles
methods, the effects of crystal field, spin-orbit coupling, and Coulomb
correlations, on comparable interaction length scales, are discussed.
While tilting induces a $t_{2g}-e_{g}$ crystal-field splitting and
band narrowing, spin-orbit coupling induces a partial splitting of
the $J_{eff}$ bands rendering SrIrO$_{3}$ a semi-metallic ground
state. The SOC enhanced hybridization of Ir-O orbitals, serve as a
explanation to why the critical Hubbard correlation strength increases
with increasing SOC strength in SrIrO$_{3}$ to induce an insulating
phase.
\end{abstract}


\submitto{\JPCM }

\maketitle

\section{Introduction}

In the Ruddlesden-Popper series Sr$_{n+1}$Ir$_{n}$O$_{3n+1}$, $n=1$
(Sr$_{2}$IrO$_{4}$) \cite{Ohsumi,Kini,Crawford,Cao,Wojek,Zocco}
and $n=2$ (Sr$_{3}$Ir$_{2}$O$_{7}$) \cite{Zocco,Bossegia,King,Subramanian,Nagai}
are insulators, while $n$ $=$ $\infty$, (SrIrO$_{3}$) is semi-metallic
\cite{Nie,Moon,Zhang-1}. All systems in this series, chemically associate
Ir with a formal valence of $+4$, due to its octahedral coordination
(IrO$_{6}$) with the neighboring ligands. The insulating ground state
of Sr$_{2}$IrO$_{4}$ and Sr$_{3}$Ir$_{2}$O$_{7}$, with Ir in
its $5d^{5}$ electronic configuration, is accounted by means of the
spin-orbit coupling (SOC)  driven $J_{eff}$ model. According to the
model, the $t_{2g}$ bands in Sr$_{2}$IrO$_{4}$ and Sr$_{3}$Ir$_{2}$O$_{7}$
are split into $J_{eff}$ = $\frac{3}{2}$ and $J_{eff}$ = $\frac{1}{2}$
multiplets \cite{Ohsumi,Cao,Fujiyama,Kim,Moon-1,Moon-2}. The $J_{eff}$
= $\frac{3}{2}$ states accommodate four electrons and therefore,
are completely filled. The remaining electron is occupied into the
$J_{eff}$ = $\frac{1}{2}$ doublet state. However, by virtue of their
narrow bandwidth Coulomb correlations set in, which splits the $J_{eff}$
= $\frac{1}{2}$ doublets into two energetically distinct Hubbard
bands, thereby, leading to an insulating ground state \cite{Pesin,Watanabe,Wang,Jackeli,Chaloupka,Carter}.

Interestingly, despite having a octahedral coordination with a formal
Ir$^{4+}$ state, SrIrO$_{3}$ has a semi-metallic ground state. Fundamentally,
this may be attributed to the underlying crystal structure, itself.
Both Sr$_{2}$IrO$_{4}$ and Sr$_{3}$Ir$_{2}$O$_{7}$ crystallizes
in tetragonal structure in which the IrO$_{6}$ motifs are stacked
along the crystallographic $c-$axis, separated by Sr ions. The IrO$_{6}$
octahedra are distorted and tilted about the c-axis by an angle $11^{o}$
in Sr$_{2}$IrO$_{4}$ and $12^{o}$ in Sr$_{3}$Ir$_{2}$O$_{7}$\cite{Subramanian,Arita,Huang}.
Since there is no direct linking of the these motifs along the $c-$axis,
both Sr$_{2}$IrO$_{4}$ and Sr$_{3}$Ir$_{2}$O$_{7}$ display a
quasi-two dimensional structure, commonly referred as single layered
and bi-layered iridates, respectively. On the other hand, SrIrO$_{3}$
crystallizes in both monoclinic and orthorhombic phases \cite{Cao-2,Longo,Zhao},
with the former being the ambient phase. The orthorhombic structure
is stabilized when synthesis conditions are adopted to high pressure
and/or when grown in thin film form on appropriate substrates \cite{Nie,Jang,Gruenewald,Zhang,Cao-3,Wu,Biswas}.
In both monoclinic and orthorhombic phases, the IrO$_{6}$ motifs
of SrIrO$_{3}$ are interlinked along all three crystallographic axes,
thereby imparting a three dimensionality to its structure. Thus, unlike
Sr$_{2}$IrO$_{4}$ and Sr$_{3}$Ir$_{2}$O$_{7}$, there exists an
additional Ir-O-Ir chemical bonding in SrIrO$_{3}$. Due to the additional
structural constraint of the ions imposed by the $c-$axial bonding
in SrIrO$_{3}$, the IrO$_{6}$ octahedral tilt is enhanced to $16^{o}$
\cite{Zhao}, partly accounting the steric displacement of the Sr
ions. 

Octahedral distortions in perovskites are generally associated with
size mismatch of the constituents and the nature of chemical bonding.
Such distortions are manifested as tilting and/or elongation of the
octahedra. Elongation usually results due to the dominant breathing
modes which originate due to charge disappropriation in the system,
as like in BaBiO$_{3}$ where Bi exists in both $+3$ and $+5$ valence
states. However, multiple valence states associated with Ir in SrIrO$_{3}$
seems less likely, and thus octahedral tilting modes appear more dominant.
Therefore, the energetics associated with the structure of SrIrO$_{3}$
can be associated with the tilting of the IrO$_{6}$ octahedra. Identifying
the IrO$_{6}$ octahedral tilt angle ($\phi$) as a physical parameter,
we study the electronic structure properties of SrIrO$_{3}$ using
first-principles density functional theory based methods. We find
that octahedral tilting not only induce a $t_{2g}$ - $e_{g}$ crystal
field splitting but also leads to significant band narrowing. Further,
one finds that inclusion of SOC in the crystal Hamiltonian effectively
splits the low energy antibonding states into $J_{eff}$ states which
are strongly hybridized. The calculations predict a semi-metallic
ground state, which are consistent with the experimental observations.
Also, following the empirical rigid band model of the alloy theory,
it is anticipated that the transport properties of SrIrO$_{3}$ would
sensitively depend on the concentration of charge carriers, as well.
Furthermore, adopting to Hubbard Hamiltonian, our calculations also
show that Coulomb correlations in SrIrO$_{3}$ have marginal effects
on its low temperature electronic structure properties.

\section{Computational details}

Optimization of the structural parameters and the electronic structure
properties of SrIrO$_{3}$ are carried out using the full potential
linearized augmented plane-wave (FP-LAPW) method as implemented in
the Wien2k code \cite{Blaha}. The LAPW sphere radii for Sr, Ir and
O were chosen as $2.26$, $2.10$ and $1.70$ a.u., respectively.
The lattice parameters were adopted to the experimental values, with
$a$$=5.591$ \AA{}, $b=$$5.561$ \AA{} and $c=7.882$ \AA{} \cite{Zhao}.
The internal ionic coordinates of the Sr and O ions, which are not
fixed by the symmetry, are relaxed using the force minimization technique.
Well-converged basis sets were ensured, by choosing the Wien2k parameters
$RK$$_{max}$= $8$ ; $G_{max}$ = $24$ and $l_{max}$ = $10$.
Additional local orbitals were also used to account for the semi-core
Ir $5p$ states. The exchange correlation potential to the crystal
Hamiltonian was considered in the generalized gradient approximation
(GGA) as prescribed by Perdew, Burke and Ernzerhof \cite{Perdew}.
The Brillouin zone $k-$mesh sampling for total energy convergence
was modeled with $384$ $k$-points in its irreducible part. The core
states were treated relativistically, while SOC was included for the
valence states through the second variational step \cite{Koelling}.

\section{Results and discussions}

\subsection{Structure}

\begin{table}[h]
\begin{tabular}{cccccccc}
\hline 
 & \multicolumn{2}{c}{Sr} & \multicolumn{2}{c}{O(1)} & \multicolumn{3}{c}{O(2)}\tabularnewline
 & $x$ & $z$ & $x$ & $z$ & $x$ & $y$ & $z$\tabularnewline
Expt & 0.0420 & 0.9928 & 0.4812 & 0.07891 & 0.7965 & 0.9602 & 0.7965\tabularnewline
Theory & 0.0422 & 0.9927 & 0.4784 & 0.08107 & 0.7950 & 0.9585 & 0.7955\tabularnewline
\hline 
\end{tabular}

\begin{tabular}{ccccc}
\hline 
 & \multicolumn{2}{c}{Bond distance ($\textrm{\AA}$)} & \multicolumn{2}{c}{Bond angles ($^{o}$)}\tabularnewline
 & Ir-O($1$) & Ir-O($2$) & Ir-O($1$)-Ir & Ir-O($2$)-Ir\tabularnewline
Expt & 2.0215 & 2.0281 & 154.205 & 152.491\tabularnewline
Theory & 2.0251 & 2.0303 & 153.346 & 152.399\tabularnewline
\hline 
\end{tabular}

\caption{\label{STRDATA}The GGA optimized structural parameters of SrIrO$_{3}$
- the ionic coordinates, selected neighboring bond distances and bond
angles, compared with the experiments \cite{Zhao}.}
\end{table}

To estimate the internal coordinates of Sr and O ions, force optimization
was performed with its tolerance set to $1$ mRy/a.u. The resulting
structural parameters are found in good agreement with the experimental
data \cite{Zhao}. The theoretical and experimental values are compared
in Table \ref{STRDATA}. 

\begin{figure}[t]
\includegraphics[scale=0.33]{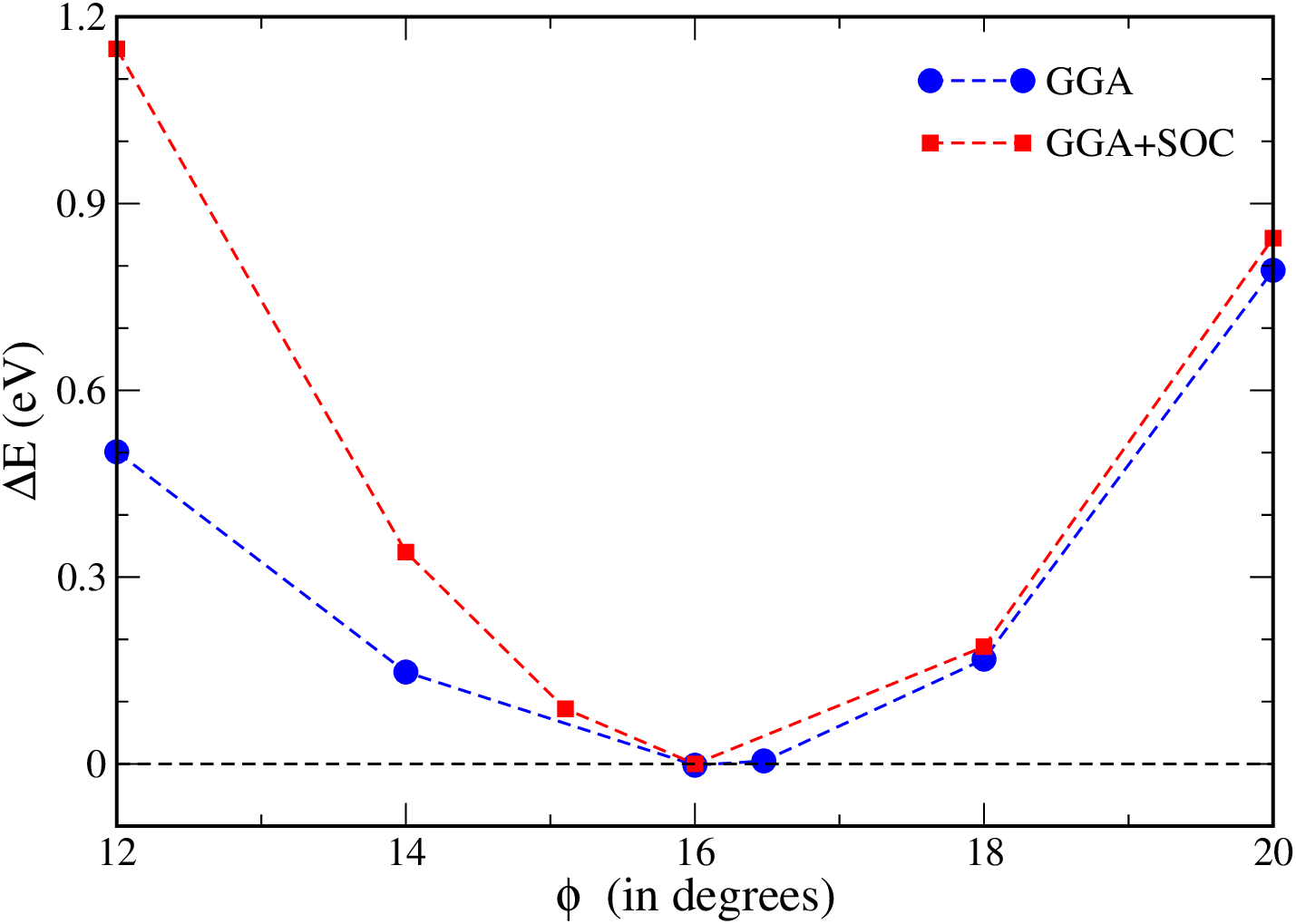}

\caption{\label{EnerPhiAngle}Color online : The relative change in the total
energy, with respect to the equilibrium, as a function of $\phi$
in SrIrO$_{3}$, calculated in the scalar relativistic GGA scheme
( blue line) and with SOC (red line) included in the Hamiltonian.}
\end{figure}

The total energy minimization were also carried out by varying $\phi$
to ascertain whether SOC has any prominent role in determining the
structural properties of SrIrO$_{3}$. The values of $\phi$ corresponding
to these non-equilibrium structures, were estimated by using the relation
$cos\theta_{1}=\frac{(2-5cos^{2}\phi_{1})}{(2+cos^{2}\phi_{1})}$
and $cos\theta_{2}$$\mbox{=}\frac{(1-4cos^{2}\phi_{2})}{2}$. Here,
$\theta_{1}$ and $\theta_{2}$ are the Ir-O(1)-Ir and Ir-O(2)-Ir
bond angles with O(1) and O(2) representing the out-of-plane and in-plane
O ions of the orthorhombic unit-cell. The average of $\phi_{1}$ and
$\phi_{2}$ ($\equiv$ $\phi$) was used as the parameter to minimize
the total energy. The variation in the total energy of SrIrO$_{3}$
with respect to tilt angle, $E(\phi)$, is shown in Fig.\ref{EnerPhiAngle}.
Compared to the experimental value of $15.8^{o}$, we find that the
scalar relativistic GGA calculations overestimate the equilibrium
$\phi$ as $16.5^{o}$, while the GGA+SOC calculations underestimate
it to $15.2^{o}$. 

The origin of octahedral tilting in perovskites is mainly attributed
to steric and/or electronic interactions. In the steric models, tilting
is facilitated due to the misfit size of the cation. Following the
GGA and GGA+SOC total energy calculations, the difference in the estimated
equilibrium value of $\phi$ differs more than $1^{o}$. Also, from
Fig.\ref{EnerPhiAngle} it follows that the stiffness of the tilting,
a quantity which is proportional to the second derivative of $E(\phi)$
curve at equilibrium, increases substantially with SOC. The estimated
increase with respect to scalar relativistic calculations is approximately
three times larger.\emph{ }Therefore, it may be conjectured from these
results that the large energy values associated with the structural
perturbations may be of electronic in origin, rather than due to steric
interaction, alone. Note that the energy associated with steric displacements
of cations in oxide perovskites are usually of the order of few meV
( $\simeq$ $10$ - $500$ K).

\subsection{Electronic structure}

\subsubsection{Scalar relativistic calculations}

\begin{figure}[t]
\includegraphics[scale=0.33]{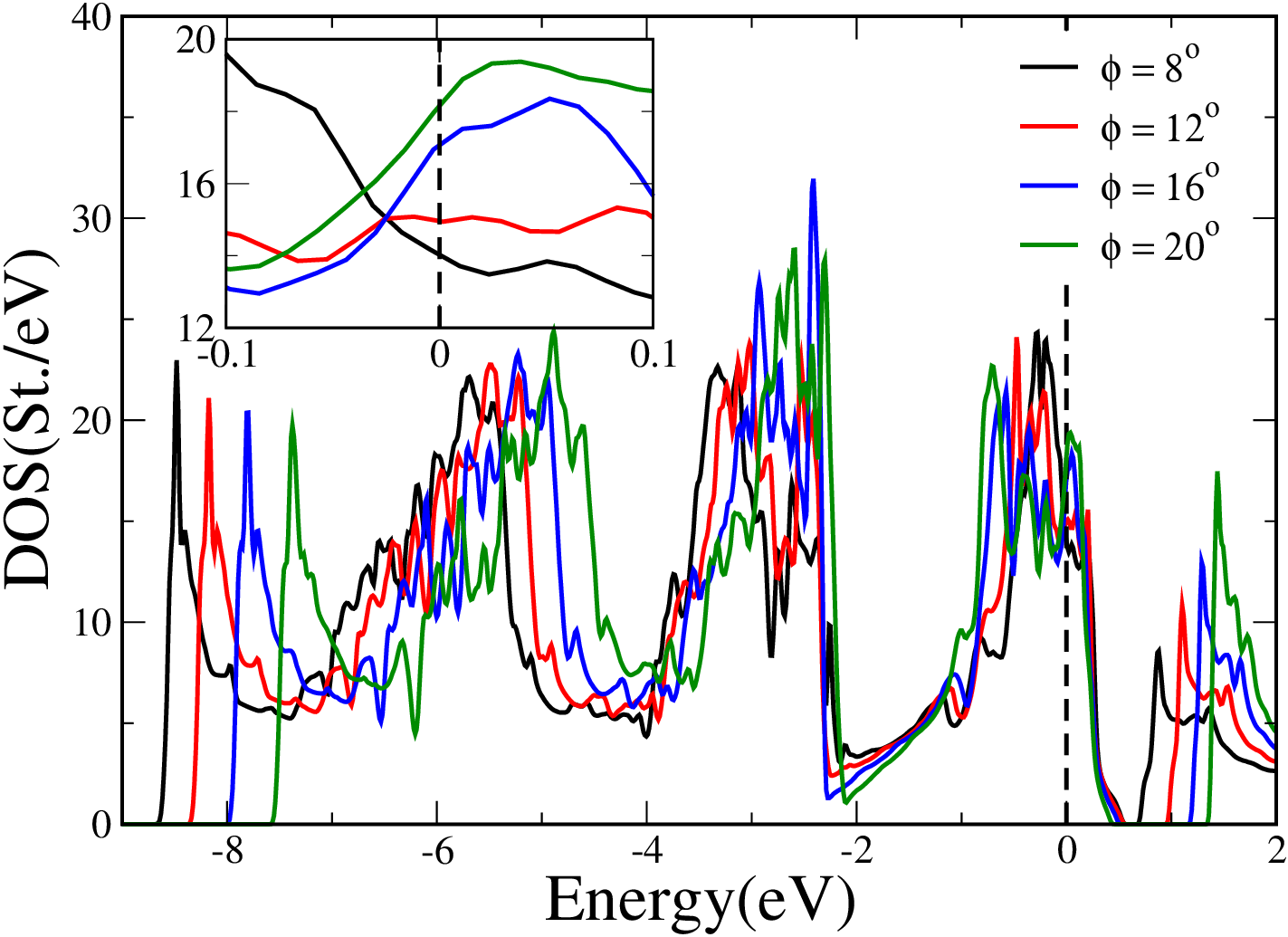}

\caption{\label{TILT_DOS_CF}Color online: The GGA derived DOS of SrIrO$_{3}$
calculated with various IrO$_{6}$ tilt angles, as indicated by the
legends. The inset shows the blow-up of the states near $E_{F}$ and
the $t_{2g}$ - $e_{g}$ crystal field splitting. The vertical broken
line (black) through energy zero represents the Fermi energy. }
\end{figure}

\begin{figure}
\includegraphics[scale=0.33]{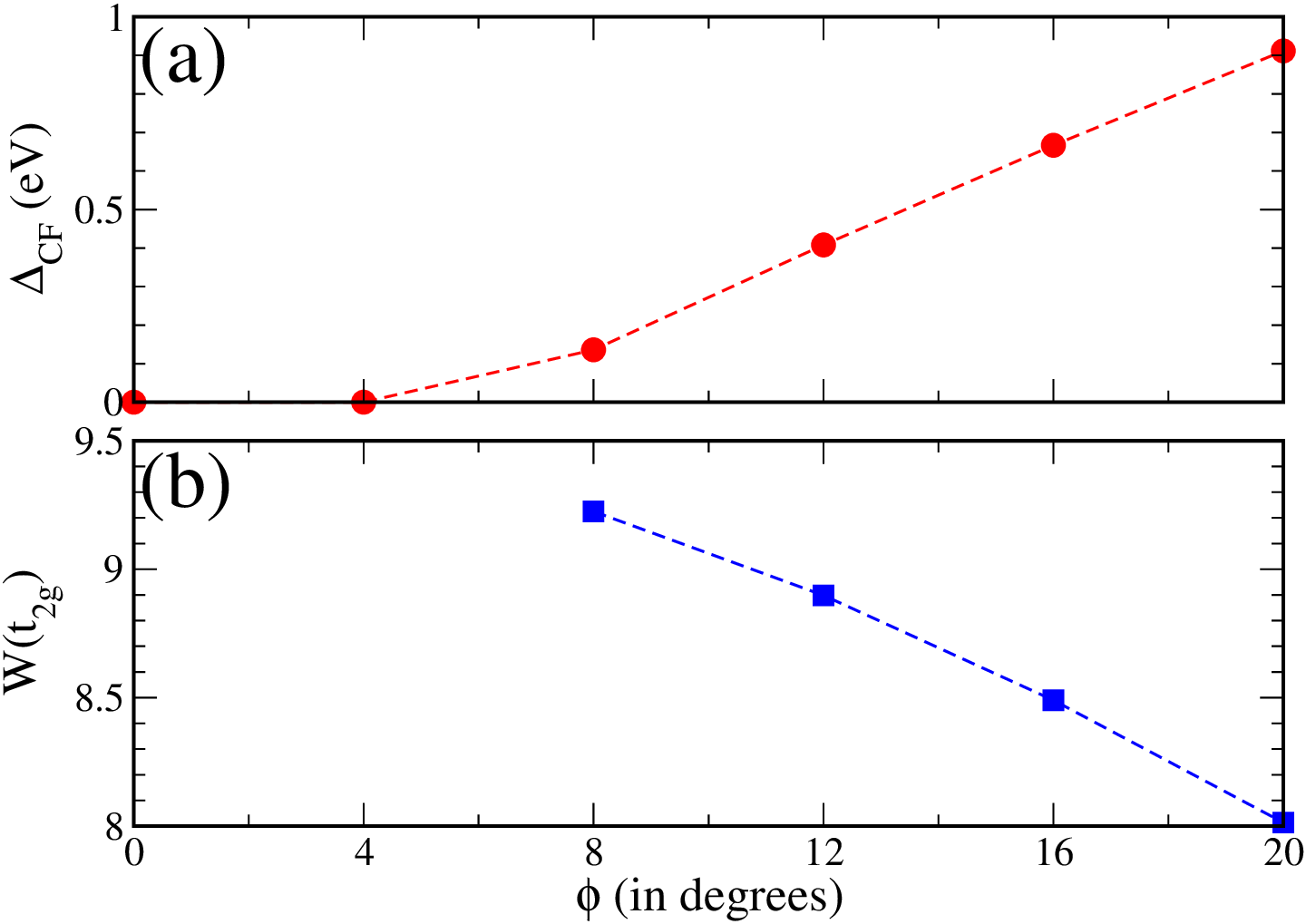}

\caption{\label{CEF_T2G}Color online: (a) The change in the magnitude of the
$t_{2g}$ - $e_{g}$ crystal field splitting of the Ir $5d$ manifold
and (b) the decreasing $t_{2g}$ band width as a function of the IrO$_{6}$
tilt angle, $\phi$ in SrIrO$_{3}$, calculated using the GGA formalism
of electronic structure method.}
\end{figure}

To study, how the crystal field splitting ($\Delta_{CF}$) of the
$t_{2g}$ - $e_{g}$ bands evolve as a function of $\phi$, we show
in Fig.\ref{TILT_DOS_CF}, the density of states (DOS) of SrIrO$_{3}$,
for $8^{o}$ $\leq$ $\phi$ $\leq$ $20^{o}$. For $\phi$ $\leq$
$4^{o}$, the $t_{2g}$ and $e_{g}$ states are strongly hybridized
and results in no electronic gap between them. A crystal field induced
gap emerges for $\phi$ $>$ $4^{o}$, the magnitude of which linearly
increases with $\phi$, as shown in the Fig.\ref{CEF_T2G}(a). The
$\Delta_{CF}$ gap resides in the unoccupied part of the DOS spectra.
The integrated DOS from $E_{F}$ to the top of the $t_{2g}$ band
is estimated as $\simeq1$$e^{-}$ per formula unit, consistent with
the $5d^{5}$ configuration of Ir ions in SrIrO$_{3}$. 

Fig.\ref{TILT_DOS_CF} also reveals significant $t_{2g}$ band narrowing
with increasing $\phi$. The bottom of the Ir $5d$ valence band at
$\simeq$ $-9$ eV below $E_{F}$ for $\phi$ $=$ $0^{o}$ linearly
shifts to lower binding energies with increasing $\phi$, and is estimated
to be at $-7.5$ eV for $\phi$ $\simeq$ $20^{o}$. Nevertheless,
the top of the $t_{2g}$ band at $\simeq$ $0.6$ eV above $E_{F}$
appears to be more or less pinned, irrespective of $\phi$, for $\phi$
$>$ $4^{o}$. The variation in the $t_{2g}$ bandwidth, $W(t_{2g})$,
as a function of $\phi$ is shown in Fig.\ref{CEF_T2G}(b). For the
equilibrium structure, the $t_{2g}$ $-$ $e_{g}$ crystal field splitting
was estimated as $\simeq$ $0.55$ eV. Thus, IrO$_{6}$ octahedra
tilting not only induce a $t_{2g}$ $-$ $e_{g}$ crystal field splitting
in SrIrO$_{3}$, but also lead to a significant $t_{2g}$ band narrowing
of the Ir $5d$ manifold.

The atom resolved partial DOS of SrIrO$_{3}$ (\emph{not shown}) also
reveal a wide spread distribution of the Ir $5d$ and O $2p$ states
over the entire energy region, suggesting a covalent nature of chemical
bonding in SrIrO$_{3}$. Both bonding and antibonding states are evident
in the valence band spectra, which are separated by a pseudo-gap like
feature at $-2.3$ eV below $E_{F}$, i.e., the states in the energy
range $-8$ $\leq$ $E$(eV) $\leq-2.3$ represent the bonding states,
while those above $-2.3$ eV constitute the antibonding states. One
may note from Fig.\ref{TILT_DOS_CF} that the relative energy position
of the bonding - antibonding states in SrIrO$_{3}$ and all non-equilibrium
structures appear to be fixed, irrespective of the IrO$_{6}$ octahedral
tilt angle.

The GGA DOS find $E_{F}$ on the higher side of the $t_{2g}$ band,
with $\simeq$ $4.25$ St./eV/f.u., indicating SrIrO$_{3}$ to be
a good metal. However, this is in sharp contrast with the experiments
which characterize SrIrO$_{3}$ as a semi-metal. Furthermore, within
the realms of Stoner theory of itinerant magnetism such high DOS at
Fermi energy, $N(E_{F})$, infers to a magnetic instability. The Stoner
product, $N(E_{F})\times I$, where $I$ is the Stoner factor is estimated
as $1.2$ in SrIrO$_{3}$, with $I$(Ir) $\simeq$ $0.574$ \cite{Garcia}.
Following these predictions from the GGA calculations, we carried
out spin polarized calculations. A ferromagnetic (FM) solution was
found with Ir local magnetic moment being $0.13$ $\mu_{B}$, which
was largely associated with the splitting of the $t_{2g}$ anti-bonding
states. Beyond, the GGA total energy difference between the non-magnetic
and ferromagnetic states was found to be nearly degenerate. The prediction
of magnetic ordering in SrIrO$_{3}$, following GGA calculations is
in contradiction with the experiments which find no long range magnetic
ordering in SrIrO$_{3}$. Furthermore, we also extended our calculations
to check for any stable antiferromagnetic solution with initial Ir
spin configuration assigned in the antiferromagnetic A-type, C-type
and G-type structures. However, all these calculations converged to
a non-magnetic solution. These results, therefore, indicate that the
scalar relativistic Hamiltonian although accounts for the crystal
field gap and $t_{2g}$ band narrowing in SrIrO$_{3}$, however, fails
to account for the distribution of electronic states, in particular,
that of the near $E_{F}$ anti-bonding states.

\subsubsection{Effects of spin-orbit coupling}

\begin{figure}
\includegraphics[scale=0.33]{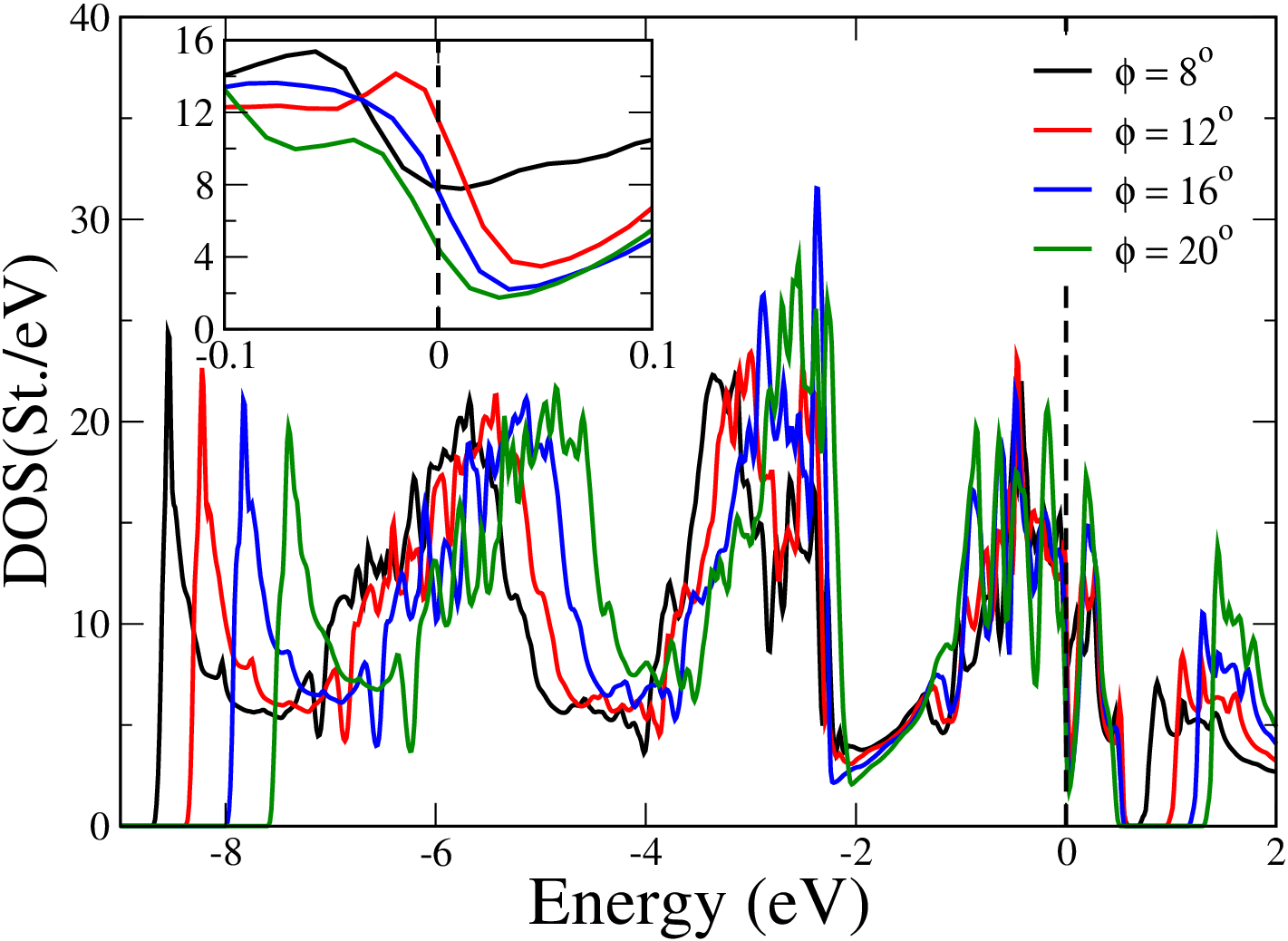}

\caption{\label{GGADOSSOC} Color online : The GGA+SOC derived DOS of SrIrO$_{3}$
as a function of IrO$_{6}$ tilt angle $\phi$, the magnitude of which
are indicated by the legends. The inset shows the blow-up of the states
near $E_{F}$. The vertical broken line through energy zero represents
the Fermi energy. }
\end{figure}

Given that SOC becomes increasingly important for $5d$ elements,
we show in Fig.\ref{GGADOSSOC}, the DOS computed by means of the
GGA+SOC calculations. The overall features of the DOS spectra in the
bonding region looks similar to that of the scalar relativistic calculations,
while the antibonding states appear to be strongly perturbed. Comparison
of Fig.\ref{TILT_DOS_CF} and Fig.\ref{GGADOSSOC} reveals that the
energy distribution of the states in the vicinity of $E_{F}$, $i.e.,$
$-0.5$ $\leq$ $E$ $\leq$ $+0.5$, undergo a drastic redistribution
with a valley like feature emerging with its depth increasing concomitantly
with $\phi$. For the equilibrium structure ($\phi$ $\simeq$ $16^{o}$)
the $N(E_{F})$ reduces to $1.88$ St./eV/f.u, which is significantly
smaller in comparison with the scalar relativistic result. The reduction
in $N(E_{F})$ is primarily due to the partial splitting of the $J_{eff}$
states of the Ir $t_{2g}$ manifold. According to the $J_{eff}$ model
of iridates, SOC splits the Ir $t_{2g}$ states into $J_{eff}$ $=$
$\frac{1}{2}$ and $J_{eff}$ $=$ $\frac{3}{2}$ states. If one associates
the Ir formal chemical valence in SrIrO$_{3}$ as $+4$, simple band
filling assuming an ionic picture, would suggest that the low energy
$J_{eff}$ $=$ $\frac{3}{2}$ states which are a linear combination
of $|m_{j}|$$=$$\frac{1}{2}$ and $|m_{j}|$$=$$\frac{3}{2}$ states
would occupy two electrons each leaving $J_{eff}$ $=$ $\frac{1}{2}$
doublet singly occupied. The integrated DOS from $E_{F}$ to the top
of the $t_{2g}$ band in SrIrO$_{3}$ yields $\simeq$ $1$ $e^{-}$per
formula unit. However, quite different from the case of insulating
Sr$_{2}$IrO$_{4}$ and Sr$_{3}$Ir$_{2}$O$_{7}$ , the $J_{eff}$
states in SrIrO$_{3}$ are found to be a superposition of both $J_{eff}$
= $\frac{1}{2}$ and $J_{eff}$ = $\frac{3}{2}$ states yielding a
semi-metallic ground state. We note that the hybridization of $J_{eff}$
= $\frac{1}{2}$ and $J_{eff}$ = $\frac{3}{2}$ states in SrIrO$_{3}$,
more or less follows from the three dimensionality of its structure,
due to which both in-plane and out-of-plane O $2p$ orbitals have
considerable overlap with the Ir $5d$ orbitals. Note that in Sr$_{2}$IrO$_{4}$
and Sr$_{3}$Ir$_{2}$O$_{7}$, there is no orbital overlap of the
Ir $5d$ orbitals with the out-of-plane O ions, primarily due to their
quasi two dimensional structure. 

We also find a strong dependence of $N(E_{F})$ with $\phi$. For
example, the $N(E_{F})$ reduces to $1.06$ St./eV/f.u for $\phi$
$=$ $20^{o}$ while, for reduced tilting $\phi$ $=$ $12{}^{o}$
it increases to $2.39$ St./eV/f.u. Apart from the decrease in $N(E_{F})$
with increasing $\phi$, we also find that the energy derivative of
DOS at $E_{F}$ also increases. The latter implies to an increase
in the effective mass or in other words, a decrease in carrier mobility
($\mu$). Overall, the GGA+SOC calculations which find a decrease,
both in $N(E_{F})$ and $\mu$ with structural perturbations clearly
suggest that the transport properties of SrIrO$_{3}$ can be controlled
by tilting the octahedra, which in turn can be accomplished by strain.
This is consistent with the strain dependent studies of orthorhombic
SrIrO$_{3}$ films grown on different substrates. Tunable semi-metallic
and metal-insulator transitions have been observed in experiments
due to lattice mismatch of the orthorhombic SrIrO$_{3}$ with that
of the substrates \cite{Gruenewald,Zhang}. It is inferred from the
experiments that in-plane compressive strain decreases the Ir $5d$
bandwidth. With the underlying symmetry of the lattice being unchanged,
it can therefore be well anticipated that the increase in the energy
of the system due to strain would be minimized mainly via the displacement
of O ions, implying IrO$_{6}$ octahedral tilting. 

Moreover, in the context that octahedral tilting in SrIrO$_{3}$ leads
to the $t_{2g}$-$e_{g}$ band splitting and narrowing of the bands,
then conversely, SOC via structural modifications enhance the extent
of Ir $5d$ and O $2p$ orbital overlap. The argument follows from
the fact that GGA when finds the equilibrium $\phi$ as $16.5^{o}$,
GGA+SOC estimates it to $15.2^{o}$. Therefore, enhanced hybridization
of Ir $5d$ - O $2p$ orbitals would reflect to an enhanced dispersion
of these hybridized bands. Based on the hybridization and Coulomb
correlations analogy, our argument is consistent with the fact that
increased SOC enhance the critical Hubbard interaction strength to
drive an insulating ground state in SrIrO$_{3}$ \cite{Zeb}.

\subsubsection{Role of carrier concentration}

Having found that the structure and spin-orbit coupling play an important
role in determining the electronic structure of SrIrO$_{3}$, we investigate
the role of carrier concentration in SrIrO$_{3}$. The blow-up of
near $E_{F}$ region, as shown in Fig.\ref{GGADOSSOC}, shows that
the position of $E_{F}$ is not at the bottom of the valley, but resides
along its negative slope. Therefore, based on the rigid band model
of electronic structure it could be anticipated that electron doping,
which can be accomplished via O defects, would slide the $E_{F}$
down the valley, thereby decreasing $N(E_{F})$ further. In general,
a low $N(E_{F})$ also indicate to higher phase stability. Therefore,
such O defects modified electronic structure appears quite favorable
in SrIrO$_{3}$, which along with structural perturbations would increase
the propensity of the system to exhibit metal-insulator transitions.

The carrier concentration in SrIrO$_{3}$ is estimated to be $\sim$
$10^{20}$ $e/cm^{3}$ \cite{Biswas}, the magnitude of which lies
in the vicinity of the semiconductor $-$ metal phase boundary of
the solid state. In the rigid band model, accommodating $10^{20}$
$e/cm^{3}$ ($\equiv$ $0.0245$ $e/u.c.$) in the DOS spectra necessitate
a shift of $E_{F}$ by $23$ meV. Such an effect significantly reduce
$N(E_{F})$ to $0.82$ St./eV/u.c., thereby increasing the propensity
of a metal-insulator transition in SrIrO$_{3}$. Moreover, for an
enhanced IrO$_{6}$ tilting, say $\phi$ = $20^{o}$, a carrier concentration
corresponding to $10^{20}$ $e/cm^{3}$ yields a of $N(E_{F})$ $0.30$
St./eV/u.c. Thus, we find that cooperative effects associated with
octahedral tilting, SOC and carrier concentration in SrIrO$_{3}$
play an important role in controlling the transport properties of
the system.

\subsubsection{Effects of Coulomb correlations}

\begin{figure}
\includegraphics[scale=0.33]{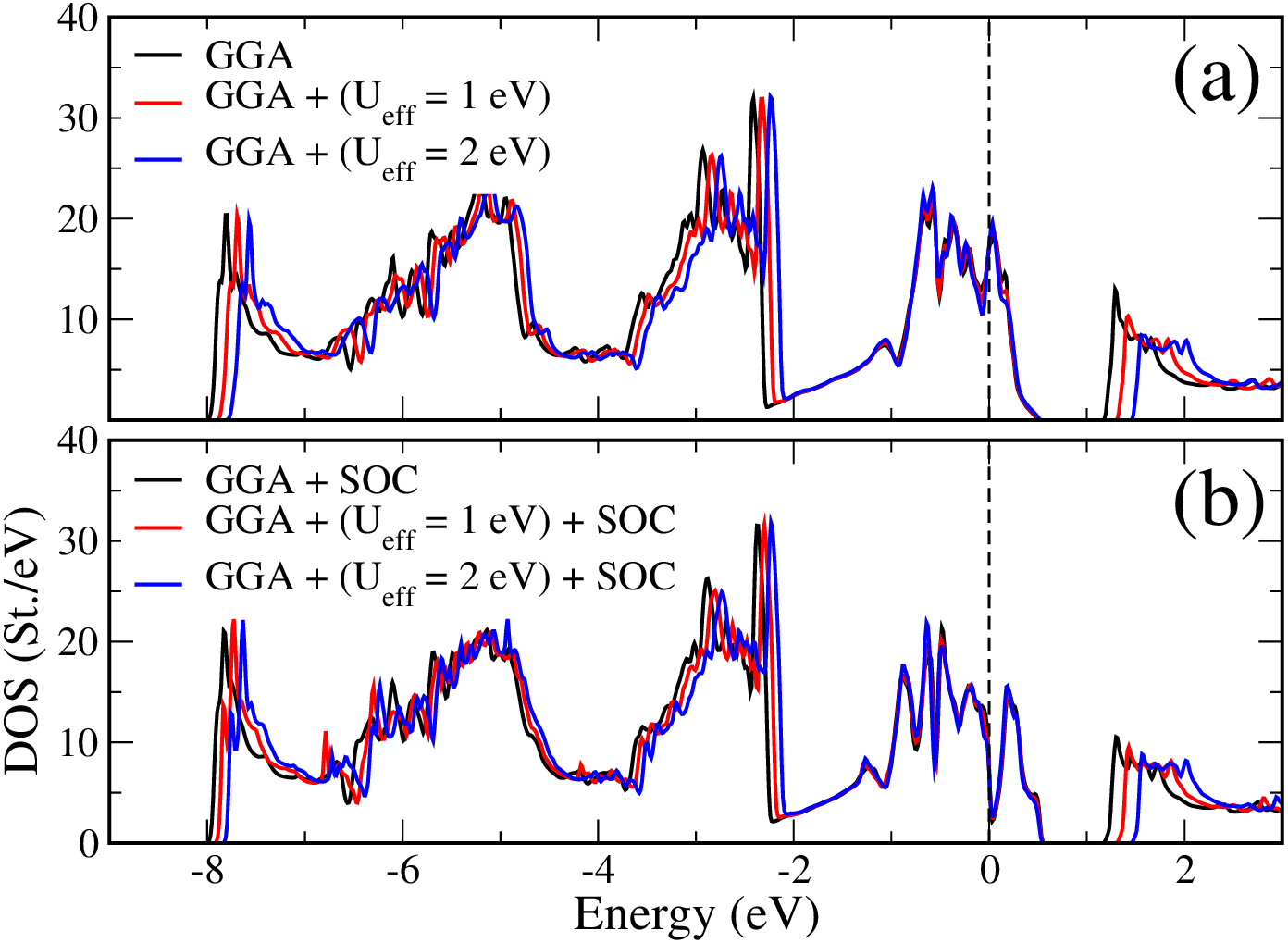}

\caption{\label{DOSGGAUSOC} Color online : The calculated DOS using the GGA+U$_{eff}$
scheme (a) without and, (b) with SOC, for the equilibrium structure
; $\phi$ $\simeq$ $16^{o}$. The vertical broken line through energy
zero represents the Fermi energy. }
\end{figure}

We also have studied the effect of Coulomb correlations on the electronic
structure of SrIrO$_{3}$. In Fig.\ref{DOSGGAUSOC}(a), we compare
the GGA and GGA+U$_{eff}$ DOS. Evidently, the most notable effect
with Coulomb correlations in SrIrO$_{3}$ is the increase in the magnitude
of the crystal field gap between the $t_{2g}$ and $e_{g}$ bands.
Also, while a small shift on the energy scale is also observed for
the bonding states, the distribution of electronic states in the antibonding
states remains more or less unchanged. On the other hand, the GGA+U$_{eff}$+SOC
calculations, the results of which are shown in Fig.\ref{DOSGGAUSOC}(b),
reveal no significant changes in the electronic structure arising
due to Coulomb correlations. Thus, on a comparative length scale ($0.5$
- $2$ eV) between the crystal field splitting, SOC and Coulomb correlations,
the effect on Coulomb correlations on the electronic structure appears
to be very marginal. 

Moreover, to study how Coulomb correlations and SOC compete in SrIrO$_{3}$,
we carried out spin polarized calculations in the GGA+U$_{eff}$ scheme
for $0$ $\leq$ U$_{eff}$ (eV) $\leq$ $4$ eV. In these calculations,
an Ir local magnetic moment was computed in the FM and AFM-A and AFM-G
structures, for all values of U$_{eff}$. For the AFM-C type ordering,
an Ir local moment appeared only when U$_{eff}$ $>$$3.5$ eV. Note
that the emergence of magnetic moment at the Ir sites with $U_{eff}$
can be attributed to the increased localization of the $5d$ states.
However, with inclusion of SOC term in the Hamiltonian, all calculations
with U$_{eff}$ $<$ $3.5$ eV, irrespective of the underlying magnetic
ordering, converged to nonmagnetic solution. Since the origin of local
moments in the GGA+U$_{eff}$ are associated with localization of
the $5d$ states, its annihilation with SOC for U$_{eff}$ $<$$3.5$
eV is to be attributed to the delocalization (or enhanced hybridization)
of the corresponding states with O $2p$ orbitals. However, we note
that a stable magnetic moment ($\simeq$ $0.5$ $\mu_{B}$) was found
for all magnetic structures in the GGA+U$_{eff}$+SOC scheme of calculations
when U$_{eff}$ $>$ $3.5$ eV, \emph{i.e., }when the magnitude of
U$_{eff}$ exceeds the Ir $5d$ bandwidth. Based on these calculations,
we argue that on a comparable length scale of interactions, Coulomb
correlations and SOC compete with each other, with SOC partly nullifying
the Ir $5d$ localization effects by enhancing the inter-atomic Ir
and O orbital hybridization.

\section{Summary and conclusion}

In summary, we find a cooperative effect on the electronic structure
of SrIrO$_{3}$ due to an interplay of its underlying lattice and
SOC. Octahedral tilting of the IrO$_{6}$ motif in SrIrO$_{3}$ appears
to be primary source of crystal field splitting of the $t_{2g}$ -
$e_{g}$ bands. The magnitude of the splitting scales with tilt angle,
which also results in band narrowing. On the other hand, SOC induces
a partial splitting of the $t_{2g}$ bands into $J_{eff}$ states
which reduce $N(E_{F})$, driving SrIrO$_{3}$ to a semi-metallic
ground state. Taking into consideration that the tilt angle can be
tuned by strain and that electron doping can be accomplished via controlled
O defects, the propensity to drive in a metal-insulator transition
in SrIrO$_{3}$ appears consistent with the experiments. Besides,
on a comparative length scale of interactions, we find that the effect
of Coulomb correlations on the electronic structure of SrIrO$_{3}$
is very nominal. Interestingly, the SOC enhanced orbital hybridization
also substantiates the preposition that the magnitude of the critical
Hubbard interaction strength to drive an insulating ground state in
SrIrO$_{3}$, increases with increasing SOC strength.

\ack{}{}

The authors thank A. K. Shukla for helpful discussions. VS acknowledges
CSIR, India for JRF fellowship and JJP acknowledges financial assistance
from the CSIR XII FYP project, AQuaRIUS.

\end{document}